\documentstyle[psfig]{mn}

\title{New binary parameters for the symbiotic recurrent nova \\
T Coronae Borealis}
\author[K. Belczy{\'n}ski and J. Miko{\l}ajewska]
       {K. Belczy{\'n}ski,\thanks{e-mail: kabel@camk.edu.pl} and  
        J. Miko{\l}ajewska \thanks{e-mail: mikolaj@camk.edu.pl}\\
        Nicolaus Copernicus Astronomical Center, Bartycka 18, 00--716 
        Warsaw, Poland}
\date{Accepted 1997 November 12.
      Received 1997 October;
      in original form 1997 April}

\pagerange{\pageref{firstpage}--\pageref{lastpage}}
\pubyear{1997}

\begin{document}

\maketitle

\label{firstpage}

\begin{abstract}
The amplitude of the ellipsoidal variability,
the mass function, and the evolutionary limits on the
component masses have been used to constrain the binary
system parameters of T Coronae Borealis. Contrary to all previous
studies, our analysis shows that the mass ratio of \mbox{T CrB}
$q \equiv M_{\rm g}/M_{\rm h} \approx 0.6$ which implies
a low mass binary system, with the stellar masses 
$M_{\rm g} \sim 0.7 \rm M_{\odot}$ for the red giant
and $M_{\rm h} \sim 1.2 \rm M_{\odot}$ for the hot companion.
This result strongly supports 
the thermonuclear runaway model  
for this recurrent nova, and solves all controversies about 
the nature of
the hot component and the physical causes of its eruptions.
\end{abstract}


\section{Introduction}

T Coronae Borealis is a recurrent nova 
which underwent major eruptions in 1866 and 1946.
Its quiescent optical spectrum shows M--type absorption
features with
the additional H\,{\sevensize I}, He\,{\sevensize I}, 
He\,{\sevensize II}, and [O\,{\sevensize III}] emission
lines, and Balmer jump (Kenyon 1986, and references therein).
Such type of optical spectrum qualified
\mbox{T CrB} to be classified as symbiotic system.
Recent classifications based on the TiO bands
in the red part of spectrum, 
indicate that the cool
component is a normal M4 III giant 
(Kenyon \& Fernandez--Castro 1987),
while the nature of its hot companion remains
controversial.

Sanford (1949), and later Kraft (1958) noted periodic 
radial velocity changes in the M giant's absorption features and
the H\,{\sevensize I} emission lines. Kraft refined Sanford's
period estimate to 227.6 days, derived a total mass of the system
of 5 $\rm M_{\odot}$, and a mass ratio of 1.4 with the giant being
the more massive component. He also noticed that the M giant
should fill its Roche lobe.
In fact, the characteristic double bump visible in 
the $VRIJ$ light curves of \mbox{T CrB} indicates that the giant 
is indeed tidally distorted (Bailey 1975; Lines, Lines \& McFaul 1988;
Yudin \& Munari 1993).
Although the observed amplitude of the light and radial velocity
changes suggests a large orbital inclination 
(Kenyon \& Garcia 1986, hereafter KG), 
the lack of eclipses in the UV continuum and emission lines
observed with the IUE indicates that the system is not eclipsing
(Selvelli, Cassatella \& Gilmozzi 1992, hereafter SCG).

Recent analysis of new radial velocity data for the giant component
in \mbox{T CrB} combined with Sanford's and Kraft's data
resulted in a new orbital solution and confirmed previous
estimates for the component masses (KG).
The spectroscopic orbit suggests that \mbox{T CrB} is relatively
massive symbiotic system, and in particular that the companion
to the M giant has a mass exceeding the Chandrasekhar limit,
and thus must be a main sequence star.
This led Webbink et al. (1987, hereafter WLTO) and
Canizzo \& Kenyon (1992) to interpretation
of the nova--like outbursts of \mbox{T CrB} in terms of 
transient phenomena in a non--stationary accretion 
disk around a main sequence star.
Unfortunately, as remarked by SCG, 
the accretion model has some weighty
difficulties when confronted with most observational data.
In fact, the mass ratio $q = 1.3$ and the resulting 
companion mass above the Chandrasekhar limit are the main  
arguments in favor of the accretion model, while
practically everything else is rather against.

SCG based on extended study of IUE spectra of \mbox{T CrB}
demonstrated that the quiescent UV characteristics of the hot component,
in particular "(1) the fact that the bulk of the luminosity
is emitted in the UV range with little or no contribution
to the optical; (2) the presence of strong He\,{\sevensize II}
and N\,{\sevensize V} emission lines, suggesting temperatures
of the order of $10^5$ K; and (3) the rotational broadening
of the high-excitation lines.", the X-ray detection,
as well as the flickering in the optical light curve
reported at several epochs, are incompatible
with the presence of main-sequence accretor, while
they find natural and physically plausible interpretation
in terms of a white dwarf acceptor.
They also discussed the spectral and photometric behavior 
of \mbox{T CrB} during the 1946 outburst,
and concluded that "(1) the spectral evolution (...) has followed
the same pattern generally observed in fast novae; (2) the 
photometric light curve has obeyed the same relation 
$M^{\rm max}_{V} - t_{3}$ followed by classical novae;
(3) the luminosity at maximum was super-Eddington, 
a distinctive signature of a TNR (thermonuclear runaway) model."
Finally, they derived the accretion rate during quiescence,
$\dot{M}_{\rm acc} \sim 2.5 \times 10^{-8} {\rm M_{\sun}\,yr^{-1}}$,
which is exactly that required by the theory to produce
a TNR every 80 years on a massive white dwarf.

Though the orbit of the M giant is now very well
established still the orbit of the companion is based
on seven $\rm H_{\beta}$ radial velocity measurements
by Kraft (1958).
The H\,{\sevensize I} emission lines in \mbox{T CrB} are however broad,
up to $\sim$ 500 km s$^{-1}$,
and affected by variable absorption,
which makes any orbital solution much more uncertain
than the formal errors quoted by Kraft may suggest
(see SCG and Warner 1995, for more detailed discussion).
Moreover, recent studies of the H\,{\sevensize I} emission line
behavior in \mbox{T CrB} have demonstrated that the H\,{\sevensize I} 
lines do not follow the orbital
motion of any of the binary components (Anupama 1997;
Miko{\l}ajewski, Tomov \& Kolev 1997).
The mass ratio, $q=1.3$, derived by KG 
from analysis of the ellipsoidal variations
in the radial velocity of the giant
does not support Kraft's result, because the authors made errors 
in their Eq.(5) (see Sec. 3.4 for details).
Thus the main argument in favor of the accretion model of \mbox{T CrB} 
outbursts does not hold any longer.  

The aim of this work was to reexamine the binary model of \mbox{T CrB}
basing on analysis of light curves and spectroscopic information. 
In particular, the ellipsoidal variability, the M giant mass
function, and the $V \sin i$ allow us to constrain the
binary parameters and to demonstrate that
the system consists of a low mass M4 giant, $M_{\rm g} \sim 0.7 \rm M_{\odot}$,
filling its Roche lobe, and a $\sim 1.2 \rm M_{\odot}$ companion
most likely a white dwarf.
Our results thus support SCG's interpretation of \mbox{T CrB}.

We describe our database in Sec. 2, analyze the data and discuss the 
results in Sec. 3,
and conclude with a brief summary in Sec. 4.

\section{Observational data}

We have collected all published photoelectric photometry of \mbox{T CrB} at
quiescent phase. For the purpose of the present study we have however
chosen only the measurements transformed to the standard Johnson's (1966)
system. 
References to our database are listed in Table 1,
the light curves are shown in Figure 1.
\begin{table}
 \caption{References to collected data.}
 \begin{tabular}{@{}lll}
 No. & Reference & Bands \\
 1 & Yudin \& Munari (1993) & {\sl J}\\    
 2 & Raikova \& Antov (1986) & {\sl U, B, V}\\
 3 & Lines et al. (1988) & {\sl U, B, V}\\
 4 & Hric et al. (1991) & {\sl U, B, V}\\
 5 & Skopal et al. (1992) & {\sl U, B, V}\\
 6 & Hric et al. (1993) & {\sl U, B, V}\\
 7 & Hric et al. (1994) & {\sl U, B, V}\\
 8 & Skopal et al. (1995) & {\sl U, B, V}
 \end{tabular}
\end{table}  
\begin{figure}
 \psfig{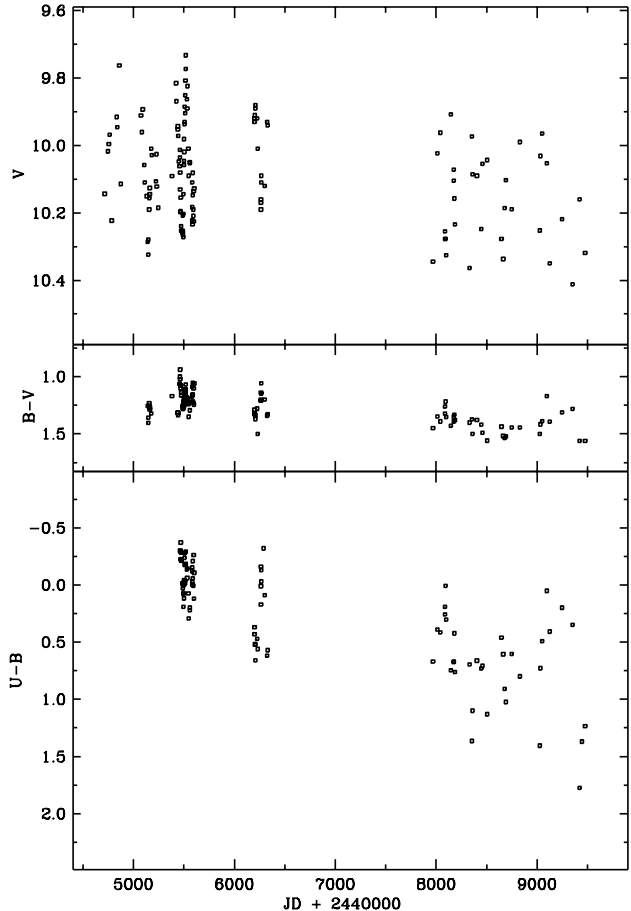}
 \caption{The $UBV$ photometry of \mbox{T CrB} in 1981--94.}
\end{figure}

The $V$ and $J$ light curves are dominated by 
sinusoidal variation with half the orbital period,
and the minima at times of spectroscopic conjunctions
are caused by orbital motion of the tidally
distorted red giant. 
Although this effect is also visible in the $B$ light,
it is superposed upon secular changes, and thus less pronounced.
The $U$ light curve is dominated by these secular changes,
as well as some erratic and perhaps quasi--periodic
variations which can be attributed to the hot component.

\section{Analysis}

\subsection{Variability}

The ellipsoidal variability of \mbox{T CrB} 
was first demonstrated by Bailey (1975).
Bailey also noticed
that for the binary parameters, $q = 1.4, i = 68^{\circ}$ (Kraft 1958;
Paczy{\'n}ski 1965a), the observed visual amplitude 
requires a very high value of the gravity--darkening coefficient
$\alpha \ga 1$.
Lines et al. (1988) derived
the amplitude of the ellipticity effect at $UBVRI$ by
Fourier analysis, and used it to find the prolateness
coefficient. They did not however attempt to refine
the parameters of \mbox{T CrB}.

Lines et al. also found additional
$\sim  55$ day variation with variable amplitude
which they attributed to semiregular pulsations of the red
giant, and suggested that the giant cannot be exactly
filling its Roche lobe at all times.
Their interpretation seems however implausible.
First, the amplitude of the $\sim 55$ day variation is 
increasing towards shorter wavelengths, while 
the M giants pulsations have the largest amplitude
in $V$ light owing to presence of strong TiO bands
in this spectral range, which is manifested
as {\it redder} $U-B$ and $B-V$ colors at maxima
than at minima -- just opposite to the behavior
observed in \mbox{T CrB}. Second, the $U-V \sim 1.1$ observed
in 1983, when the variation had the largest amplitude,
is much lower than $U-V \sim 3.4$ expected for
M3--4 III giants (Straizys 1992)
which suggest very low contribution of
the M giant to the $U$ light. Finally, the $V$ amplitude
of this additional variability was largest in 1983
when the hot component was bright in the optical range
as indicated by flickering observed in $B$ and $V$ light
(Lines et al. 1988). 
All this points to the hot component as the source
of $\sim 55$ day variation, although its origin is not clear.

Yudin \& Munari (1993) published $J$ light curve
of \mbox{T CrB} based on 5.8 orbital cycles, and did not find
any evidence for the M giant changing its intrinsic brightness
by more than a few hundredths of magnitude.
Their results provides additional strong support
that the erratic and quasi--periodic large amplitude
variations reported in the optical must be related
to the M giant's companion.

The secular trends in the light curve of \mbox{T CrB} have been
recently studied by Leibowitz, Ofek \& Mattei (1996).
Using amateur astronomers' visual observations 
spanning a period of nearly 40 years, they 
found a quasi--periodic, $\sim 27$ yr, oscillation superposed
on a linear fading with an average rate of 
$\sim 10^{-5}$ mag/day.
The interval covered by their data sample
is however only 50\% longer than
the estimated period, which 
combined with rather large uncertainties in the visual magnitude
estimates by various observers makes that result debatable.

Nevertheless, the occurrence of high and low luminosity
states of the hot component is also suggested by the IUE observations (SCG),
and optical emission line behavior 
(Anupama 1997; Miko{\l}ajewski et al. 1997).
>From the data in Figure 1 
we estimate the magnitude decrease of $\sim 0.09$ mag
in $V$, $\sim 0.24$ mag in $B$, and $\sim 1.1$ mag in $U$, respectively,
during the 1981--1994 period.
The IUE observations reported by SCG also suggest that 
the hot component was apparently in much higher luminosity
state in 1981--85 than in the 1990's.
In 1996, the hot component regained the brightness
level from early 1980's (Hric et al. 1997; 
Miko{\l}ajewski et al. 1997).

In our analysis we will focus on the $V$ and $J$ light curves
where the M giant provides the dominant contribution,
and the ellipsoidal variability is easily seen.
Because of noticeable decrease of brightness in $V$ 
the data have been divided into two subsets. 
First subset contains points from the left part of data
(JD 2444500--2446500) and second from the right part of it (JD
2448000--2449500) -- see Fig. 1.
The both subsets of data in $V$ and the set of $J$ data were phased and are 
shown in Figs 2 and 3, respectively. 
The ephemeris of Lines et al. (1988): 
\begin{equation}
  {\rm Min\ I}={\rm JD}\ 2431931.05 + 227.67\ E,
\label{eqmf}
\end{equation}
was used to phase the data. The initial epoch is a time of 
spectroscopic conjunction with the M giant in front.

\subsection{Admissible parameters of the \mbox{T CrB} binary system}

Since \mbox{T CrB} is noneclipsing system, the ellipsoidal light variation
and the spectroscopic orbital solution for the M giant
are not sufficient to fully constrain the system parameters
(Morris 1985; Hall 1990).
We need additional constraints. 

\begin{figure}
 \psfig{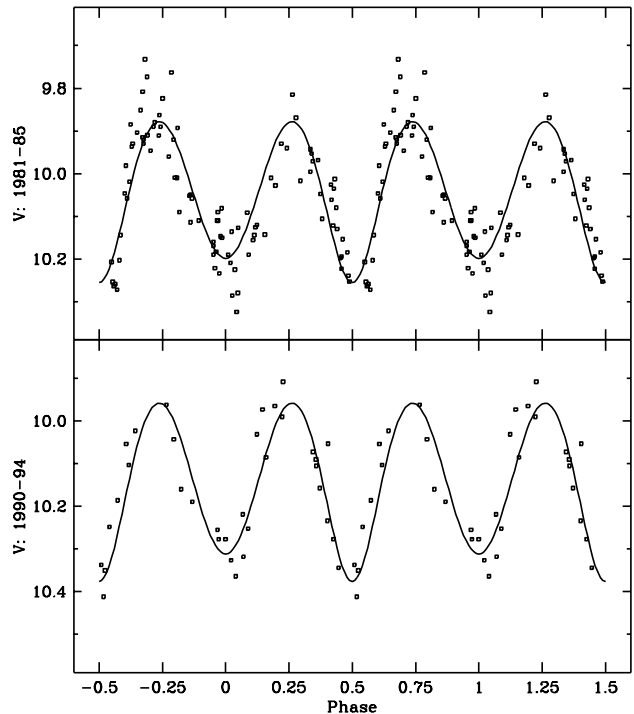}
 \caption{Synthetic $V$ light curves for the model with
   $q=0.6, i=60^{\circ}, \alpha=0.95$ and $e=0.0$.}
\end{figure}

\begin{figure}
 \psfig{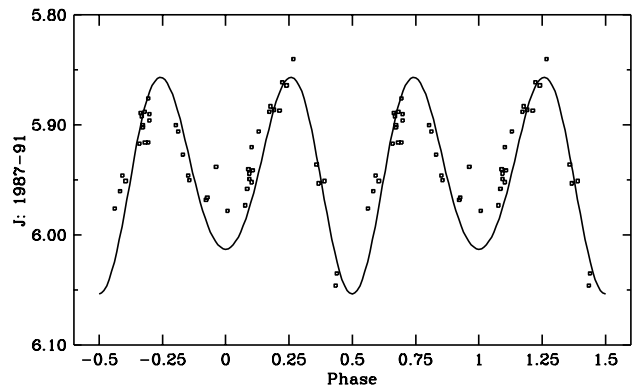}
 \caption{Synthetic $J$ light curve for the model with 
$q=0.6, i=60^{\circ}, \alpha=0.32$ and $e=0.0$.}
\end{figure}

As demonstrated by SCG, the quiescent IUE observations
combined with X-ray detections, flickering,
and the outburst behavior of \mbox{T CrB} point out toward
a massive white dwarf as the hot component in the system. 
The only problem met was the hot component's radial velocity 
measurements by Kraft (1958), which resulted in its mass
above the Chandrasekhar limit. 
To solve this problem, SCG just stretched the probable errors
of Kraft's measurements to show that the massive white dwarf
can be compatible with his solution.
Basing on the arguments given by SCG (recalled here in Sec. 1),
and the fact that recent studies have shown that the 
Balmer H\,{\sevensize I} emission lines in \mbox{T CrB}
do not follow the hot component (see also Sec. 1),
we reject the orbital solution for the hot component,
and instead we assume that its mass does not
exceed the Chandrasekhar limit, $M_{\rm h} \la 1.44 \rm M_{\odot}$.
We also assume that a reasonable limit to the red giant mass is 
$M_{\rm g} \ga 0.6 \rm M_{\odot}$, which ensures the secondary can
evolve to giant dimensions during the lifetime of the Galaxy
(e.g. Webbink 1988).

\begin{figure}
 \psfig{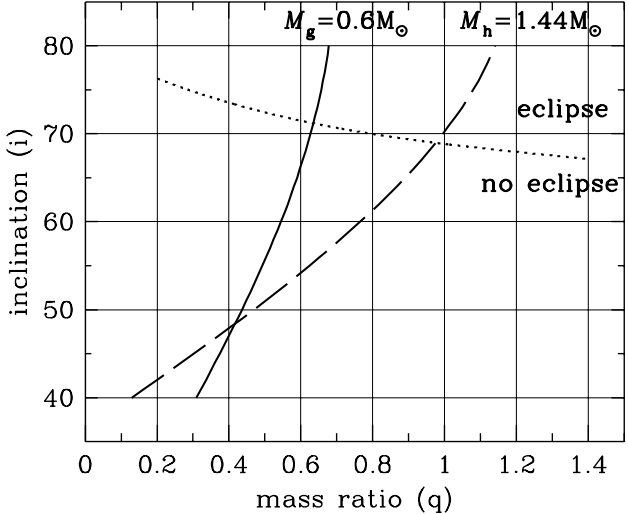}
 \caption {Parameter space diagram for \mbox{T CrB}.
           The permitted region is confined by the line of eclipses
           (dotted curve), and two constraints for the components' masses:
            $M_{\rm h} < 1.44 {\rm M}_{\odot}$ (dashed curve) and            
            $M_{\rm g} > 0.6 {\rm M}_{\odot}$ (solid curve).}
\end{figure}

Fig. 4 presents $i$ versus $q$ diagram for \mbox{T CrB}.
Using the mass function derived by KG
\begin{equation}
f(M) = M_{\rm h}\ \sin^{3}i/(1+q)^{2} =
0.30 \pm 0.01\ [\rm M_{\odot}]
\label{eqmf}
\end{equation}
with $q=M_{\rm g}/M_{\rm h}$, we plotted the 
lines corresponding to $M_{\rm h} = 1.44 {\rm M_{\odot}}$ with \mbox{T CrB} 
being to the left,
and $M_{\rm g} = 0.6 {\rm M_{\odot}}$ -- with \mbox{T CrB} to the right, respectively.
The line of eclipses, with \mbox{T CrB} being below it,
further constraints the possible values of $q$ and $i$.

Figure 2 shows that any reasonable mass ratio
is below 1.0,
which implies that the red giant 
is the less massive component of \mbox{T CrB}!

The low mass ratio is also in better agreement with
the relatively low rotational velocity,
$V_{\rm rot} \sin i \la 10$ km/s, reported by KG.
Since the giant cannot rotate faster than synchronously
with the orbit, the rotational velocity provides 
the lower limit for $q$. KG estimated $q_{\rm min} = 0.4$.
In \mbox{T CrB}, where ellipsoidal light variations 
demonstrate the importance of tidal effects, the giant's
rotation should be synchronized with the orbital motion.
Zahn (1977) derived synchronization and circularization
time scales for convective stars in good agreement
with the observations of binary systems.
Using his Eqs. (61) and (62) we  estimate
$t_{\rm synchr} \sim 300$ yr, and $t_{\rm circ} \sim 3000$ yr,
respectively, for $q \la 1$.
KG noticed that the giant would be not synchronized
if it evolved up to the giant branch on a rapid timescale
($\la t_{\rm synchr}$). It does not however seem very plausible.
They also remarked that limb darkening and radiation
from the hot component can reduce the observed
$V_{\rm rot} \sin i$. 
Both effects
can be important in the case of \mbox{T CrB}. The radial velocity
measurements used by KG come from the period 1982-85,
when the hot component was relatively bright, and
its contribution to the light the 5200 ${\rm \AA}$ bandpass
was more than 10 \% as indicated by flickering
in $V$ and $B$ light observed in 1983 by Lines et al. (1988). 
We believe that the combined effect of additional hot
component radiation and limb darkening can easily reduce
the observed $V_{\rm rot} \sin i$ by 15-30 \% with
respect to the true value,
and raise the mass ratio to $q \sim 0.5-0.6$.
It is however unlikely, to increase by that means 
$V_{\rm rot} \sin i$ by a factor of $\sim 2$,
and make it compatible with $q=1.3$.

\subsection{Light curve synthesis}

Synthetic light curves have been computed
using Wilson--Devinney code (Wilson 1990, 1992)
for the admitted range of the mass ratio, $q=M_{\rm g}/M_{\rm h}$,
and the orbital inclination, $i$ (see Figure 2).
Models were calculated for semi--detached configuration
with the hot component very small as compared to the
Roche lobe filling M giant.

Linear limb--darkening law has been assumed,
and the $x^{V}=0.95,\ x^{R}=0.8,\ x^{I}=0.6$
and $x^{J}=0.5$ coefficients
have been adopted in the $V$ ($\lambda_{\rm eff}=5500 
{\rm \AA}$), $R$ ($\lambda_{\rm eff}=7000 {\rm \AA}$),
$I$ ($\lambda_{\rm eff}=8800 {\rm \AA}$) and 
$J$ ($\lambda_{\rm eff}=12500 {\rm \AA}$),
respectively. These coefficients have been interpolated
from the tables of Van Hamme (1993),
for the temperature $T_{\rm eff}(\rm M4III)=3560$ K
(Ridgeway et al. 1980).
The gravity--darkening exponent, $\alpha$, 
defined through dependence of luminosity on local 
surface gravity, $L \sim g^{\alpha}$,
has been taken as free parameter 
so long as $0.32 \leq \alpha \leq 1.0$,
the theoretical values for stars with convective (Lucy 1967)
and radiative envelopes (Von Zeipel 1924a,b,c), respectively.
The black body approximation for wavelength dependence
has been assumed in all our computations.
 
The IUE and optical spectrophotometry (SGC, WLTO, KG) suggests the
hot component contributes very little to the total light of
system beyond $\sim 5000 {\rm \AA}$, to become practically invisible 
in the infrared (SCG, WLTO, KG). 
We have thus attributed all light in $J$ passband 
to the red giant. 
The observed values of the $B-V$, and $U-B$ colors 
(Fig. 1) indicate very low contribution of the hot component
to the $U$ light, and none observable contribution
to the $B$ and $V$ light in the period 1990-94. 
The optical faintness of the hot component at that epoch is also suggested
by the appearance of the optical spectrum -- the continuum and absorption 
features of the M4 giant with very faint H\,{\sevensize I} Balmer
emissions, and the lack of any
flickering variability (Dobrzycka, Kenyon \& Milone 1996).
The presence of flickering (Lines et al. 1988) and emission
lines in the optical spectrum (KG), as well as $B-V$ and $U-B$ 
colors observed in 1981--85 suggests some hot component
contribution to the $V$ light at that period.
Comparing the average $V$ magnitudes for \mbox{T CrB} in these
two periods, and assuming that the contribution
of the hot component to the total $V$ light was negligible
in 1991--94, we estimate that this contribution in period 
1981-85 was about 10 per cent.
We have also neglected the hot component contribution to 
the $J$ light for the whole analyzed period.
Based on the IUE data from 1979--90 
SCG demonstrated that the hot component 
luminosity never exceeded 100 $\rm L_{\odot}$,
thus any reflection effect produced by the hot component 
in the red giant is negligibly small for the whole period.

The ellipsoidal variation has much larger amplitude
in $V$ than in $J$ band: $\Delta V \sim 0.4$, and $\Delta  J
\sim 0.15$, respectively. So different amplitudes cannot be
reproduced by our model light curves for any set of parameters.
In particular, the large amplitude in $V$ light requires 
generally large values of the gravity--darkening exponent,
$\alpha \sim 1$ for any binary parameters, while the much
lower $J$ amplitude is consistent with $\alpha \sim 0.32$
for any reasonable $q$ and $i$. The need of very high value
of $\alpha \ga 1$ to account for the observed visual
amplitude was already reported by Bailey (1975), who adopted 
the Kraft mass ratio, $q$ = 1.4, and the highest possible
inclination $i$ =$68^{\circ}$ (for which the system is still 
not eclipsing).

In this situation, we have analyzed the $V$ and $J$ light curves
separately. The grid of $V$ and $J$ light curves was generated  
for $q$ and $i$ in the range suggested by Figure 2, 
and $\alpha$ as a free 
parameter. The synthetic light curves fit
the observations fairly well in the range 
$0.4 < q < 0.8$ and $55\degr < i < 65\degr$,
with $\alpha \sim 1$ for the $V$ light curve
and $\sim 0.32$ for the $J$ light curve.
Figs 2, and Fig. 3 present examples of our solutions for 
$V$ and $J$, respectively. Identical light curves are 
obtained for the other admitted values of $q$ and $i$.
Below we use the middle values, $q= 0.6$ and $i= 60\degr$,
with the errors defined by their extremes.
These errors propagate onto our results (see Table 2).
Our solutions are presented in Figs 2 and 3. 

For mass ratios $q < 1$ adopted in our study
the red giant is more tidally distorted 
then in the models with $q \geq 1.0$,
which implies generally larger amplitudes of ellipsoidal variation,
and allows the gravity darkening exponent $\alpha$ to be smaller  
in better agreement with the theory. Unfortunately, the inconsistency
between the $V$ and $J$ amplitudes of our synthetic light curves
remains.
In particular, this inconsistency 
cannot be explained by poor quality of the $J$
light curve. Using the values of parameters derived from
the $V$ light curves we estimate the amplitude 
of the ellipsoidal variation, $\Delta J_{\rm I} = 0.20$,
and $\Delta J_{\rm II} = 0.29$, for the primary and secondary minimum,
respectively. These values exceed the observed amplitudes
by $\sim 0.1$, which is much more than accuracy of the observations 
in the $J$ band. According to Yudin \& Munari (1993) the J data
have internal accuracy better than 0.02 mag, and the data from
different orbital cycles fit the same light curve 
with similar accuracy. 

To find out why we get so different values of 
the gravity darkening coefficient for different light curves,
we have calculated the amplitudes in the $R$ and $I$ bands
for $q=0.6$, $i=60^{\circ}$, and two
values of $\alpha$: 0.32, and 1, respectively.
Our calculated $\Delta I_{\rm I} = 0.17$,
$\Delta I_{\rm II} = 0.24$ ($\alpha = 0.32$),
well agree with the observed amplitude $\Delta i \approx 0.17$
reported by Lines et al. (1988), and $\Delta I \approx 0.2$
estimated from the light curve published by Miko{\l}ajewski
et al. (1997), while the observed amplitudes in the red band,
$\Delta r \approx 0.3$ (Lines et al. 1988), and
$\Delta R \approx 0.33$ (Miko{\l}ajewski et al. 1997)
are in agreement with the model values, $\Delta R_{\rm I} = 0.29$,
$\Delta R_{\rm II} = 0.36$, for $\alpha = 1$.
Thus we meet again the problem
that the ellipsoidal variability in $I$ light 
can be reproduced with the lower (convective) value of 
$\alpha = 0.32$, while the observed amplitude in $R$ band
requires high (radiative) values of $\alpha \sim 1$. 

We believe the high values of $\alpha$ suggested by visual 
and red amplitudes
of the ellipsoidal variation result from the black body 
approximation for wavelength dependence in 
the Wilson--Devinney code. 
This assumption is probably not valid in the case
of M type stars with strong TiO bands in the optical and red part
of spectrum. 
The TiO bands are very
sensitive to even small changes in the effective temperatures,
and so can strongly affect the broadband $V$ and $R$ magnitudes
giving rise to much larger light changes than calculated
under black body assumption in the WD code.
For $q$ = 0.6, $i = 60^{\circ}$, and $\alpha = 0.32$, we estimate
the mean gravity ratio $g_{\rm min I}/g_{\rm max} \sim$ 0.84 
(where $g_{\rm min I}$ and $g_{\rm max}$ have been averaged
over the M giant's surface visible at minimum and maximum,
respectively), and accordingly averaged effective 
temperatures, $T_{\rm min I} \sim$ 3500 K, and $T_{\rm max} \sim$ 3560 K,
respectively. Adopting $(V-J)_{\rm min I}$ = 4.45, and  $(V-J)_{\rm max}$
= 4.19 (M4.3 III; 3500 K, and M4 III; 3550 K, respectively; Straizys 1992),
the $J$ amplitude, $\Delta J_{\rm I} = 0.15$, implies 
$\Delta V_{\rm I}$ = 0.41. The later value is very close to the
observed $V$ amplitude, while the $V$ amplitude calculated
in the WD code for the same parameter set, $\Delta V_{\rm I}
= 0.21$, is by a factor of 2 lower than the observed one!

There is a moral in that to analyze the ellipsoidal
variability in symbiotic binary systems with M giant components,
either one should base the analysis on the infrared light curves
(where the black body approximation for wavelength dependence
is acceptable) or, if only the optical data are available, 
model M giant atmospheres should be used for the wavelength 
dependence.

Table 2 summarize the adopted parameters for \mbox{T CrB}.
Assuming that the M giant fills its tidal lobe,
we have estimated its radius and luminosity,
and the distance to \mbox{T CrB}.

\begin{table}
 \caption{Adopted parameters for \mbox{T CrB}.}
 \begin{tabular}{@{}lc}
 Parameter & Value \\
Mass ratio, $q$ & 0.6 $\pm 0.2$ \\
Orbital inclination, $i$ & $60\degr \pm 5\degr$ \\
Gravity--darkening, $\alpha$ & 0.32 \\
Hot component mass, $M_{\rm h}$ & $1.2 \pm 0.2 \rm M_{\odot}$ \\
M giant mass, $M_{\rm g}$ & $ 0.7 \pm 0.2 \rm M_{\odot}$ \\
Orbital separation, $a$ & $0.9 \pm 0.1$ au  ($194 \pm 22 \rm R_{\odot}$) \\
M giant radius, $R_{\rm g}$ 
& $0.34 \pm 0.02$ au ($66 \pm 11 \rm R_{\odot}$) \\ 
M giant temperature, $T_{\rm eff}$ & 3560 K \\
M giant luminosity, $L_{\rm g}$ & $620 \pm 120 \rm L_{\odot}$ \\
Distance, $d$ & $960 \pm 150$ pc \\
 \end{tabular}
\end{table}

\subsection{Spurious eccentricity induced by tidal distortion}

So far our analysis has been made under assumption of
circular orbit of the \mbox{T CrB} system.
KG however noted that an eccentric orbit 
with $e=0.012 \pm 0.005$ slightly improves the fit
to the radial velocity data. 
They interpreted that eccentricity as resulting from 
the contribution of the axial rotation of the tidally deformed giant,
which has a nonuniform surface brightness, to its observed radial velocity.
The effect was studied in detail by Sterne (1941), who demonstrated
that it gives rise to a spurious eccentricity, $e_{\rm t}$,
in the orbital solution given by
\begin{equation}
  e_{\rm t}=1.5\ q^{-1}(1+q)(R_{\rm g}/a)^4 \sin i\ f(x, \beta_2),
\label{eqmf}
\end{equation}
where $R_{\rm g}$ is the giant's radius, $a$ is the orbital separation,
and  $f$ is the function of the selective gravity--darkening 
coefficient $\beta_2$, and the limb--darkening coefficient $x$:
\begin{equation}
   f(x, \beta_2)=\frac{8 \beta_2 - 3x \beta_2 - 5x}{20(3-x)}.
\label{eqmf}
\end{equation}   
The coefficient $\beta_2$ can be estimated from
\begin{equation}
   \beta_2 = \alpha \frac{1.43879 \times 10^8/\lambda T}
   {1 - \exp(-1.43879 \times 10^8/\lambda T)}
\label{eqmf}
\end{equation}
(the gray body approximation).  
Sterne (1941) has also shown that for 
a circular orbit the spurious longitude
of periastron $\omega_{\rm t} = 90^{\circ}$ or $270^{\circ}$
according to whether $f$ is positive or negative.
KG derived $\omega=80^{\circ} \pm 6^{\circ}$, which 
suggests the tidal distortion is the dominant source
of the eccentricity they found.
KG has also proposed to use that eccentricity as an
indirect measure of the mass ratio. Unfortunately,
their Eq. (5), as well as their Fig. 5, 
used for that purpose contain errors, and
their value of $q=1.3$ is wrong.

Using our expression for $e_{\rm t}$,
we find very weak dependence of $q_{\rm min}$ (for 
a lobe--filling giant) on the spurious eccentricity
$e_{\rm t}$. In particular, the term $q^{-1}(1+q)(R_{\rm g}/a)^4$
changes from 0.032 to 0.056 for $q$ increasing from
0.5 to 2.0. 
Adopting reasonable values of $x \sim 0.9$,
and $\beta_2 \sim 2.3$ (the value corresponding
to $\lambda \sim 5200 \AA$, $T = 3560 {\rm K}$,
and $\alpha = 0.32$), $q = 0.6$
and $i = 60^{\circ}$,
we estimate $f(0.9, 2.3) = 0.18$, and 
$e_{\rm t} = 0.009 \pm 0.002$ which is very close to 
$e=0.012 \pm 0.005$ derived by KG. Although this is a very
rough estimate due to crudity of the adopted values
upon which it depends, it strongly points to 
tidal effects as the source of the
eccentricity reported by KG.
Moreover, this results also provides significant
support for the low value of the gravity--darkening
exponent, $\alpha = 0.32$. Higher $\alpha$'s result
in higher values of $f$, and so $e_{\rm t}$. For example,
$\alpha=1$ will increase our $e_{\rm t}$ by a factor of 4. 

\subsection{Asymmetry in the orbital light curve}

In addition to erratic and secular variations caused by
the hot component, the orbital V light curves of \mbox{T CrB} (Fig. 3) 
show some systematic asymmetry: the ingress to the primary minimum 
is slightly longer than the egress, and in the 1981--1985 period
the maximum following the primary minimum (Max I; $\phi \sim 0.25$) 
seems to be lower than the second maximum (Max II; $\phi \sim 0.75$).
It is hard to say whether this asymmetry is also present
in the $J$ and the 1990--94 
$V$ light curves because there is not enough data points. 

There are many possible causes of asymmetry in orbital light
curves: noncircular orbit, hot or cool spots on the cool giant, 
asymmetry in the hot component.

\subsubsection{Eccentric orbit}

To check whether an eccentric orbit can 
account for the asymmetric shape of the V light curve,
we have calculated synthetic light curves for fixed
values of $q=0.6$, $i=60^{\circ}$,
and $\alpha=0.95$ (our best solution for $V$ light curves
from Sec. 3.3), with $e$ and $\omega$ as free parameters.
The model with $e=0.05$ and $\omega=120^{\circ}$
(measured from the ascending node as in Sterne (1941) and KG)
reproduces reasonably well the distorted shape of the primary
minimum and the difference in heights of maxima.
The required value of $e$ exceeds by more than 
$3 \sigma$ the eccentricity $e = 0.012 \pm 0.005$ found by KG.
The comparison of this result with the eccentric orbit
derived by KG is however not so straightforward.
If the orbit is indeed eccentric,
one should expect the spurious tidal eccentricity
to combine with the real $e$, to yield a resultant
(which is the spectroscopically measured one)
which can be either larger or smaller than the real $e$.
So, the radial velocities should be corrected
for the tidal effects before solving for the spectroscopic 
orbital elements.
As we have demonstrated in Sec. 3.4, the spectroscopic
eccentricity found by KG can be fully accounted by
the tidal distortion of the M giant.
Thus there is no evidence for any real eccentricity
in the radial velocity data.

\subsubsection{Asynchronous rotation and reflection}

Leibowitz et al. (1996) recently analyzed visual magnitude
estimates of \mbox{T CrB} spanning a 40-year period, and 
found that the photometric minima are systematically
delayed (the primary minimum by $4 \fd 7$, and the secondary by
$1 \fd 7$, respectively) with respect to the times of spectroscopic 
conjunctions given by KG. They argued that this effect is due
to asynchronous rotation of the M giant. 
If the giant rotates slower than synchronously
the tidal distortion wave on its surface is 
lagging behind the interbinary radius vector
(Lecar, Wheeler \& McKee 1976), and the ellipsoidal
light minima will be delayed with respect to spectroscopic conjunctions,
but there will be no difference in the delay times of the two minima.
Leibowitz et al. proposed
that the difference in the lags of the two minima,
and the general asymmetry in the orbital light curve
is caused by combined effects of the giant's asynchronous rotation
and the illumination of the giant's atmosphere by the hot
companion. 
Their model however faces serious problems when confronted
with the observational data.

First of all, there is no observational evidence for
significant reflection effect in \mbox{T CrB}. The hot component 
is not very luminous (SCG, and Sec. 3.3), while the TiO
band depths do not show any measurable phase dependence
(Kenyon \& Fernandez--Castro 1987)
indicating that any temperature contrast on the giant's
surface, $\Delta T_{\rm eff} \la 50 \rm K$.
Moreover, even if there is any reflection effect, 
our exemplary synthetic light curves show 
the primary maximum ($\phi = 0.25$) to be higher than 
the secondary ($\phi = 0.75$), and the interval between
the primary and secondary minima should be larger
than 0.5 $P$, contrary to what we do observe in \mbox{T CrB}.

Finally, as we argue in Sec. 3.2, for the mass ratio
resulting from our analysis, much lower than any previous
estimates, the low rotation velocity reported by KG does not 
necessarily imply that the rotation of the giant is not
synchronized with the orbital rotation.

\subsubsection{Accretion disk with asymmetric brightness distribution}

Analysis of the slope and intensity of the IUE continuum 
led SCG to the conclusion, that the bulk of the UV luminosity
of \mbox{T CrB} originates from a nonstationary accretion disk around
a white dwarf. They  also remarked that though
the disk luminosity contributes mostly to the satellite UV,
there should be also some disk contribution (a few $\rm L_{\odot}$)
to the optical luminosity of \mbox{T CrB}. This contribution
is clearly visible in the $U,B$ and $V$ light in 1981--85, while
practically absent in 1990--94.
Comparison of the average $U,B$ and $V$ magnitudes in these
two periods indicates the optical luminosity
of the disk was $L_{UBV} \ga 7 \rm L_{\odot}$ in 1981--85,
in agreement with the value predicted by SCG.
The data in Table 1 of SCG also indicate that in 1989 the average
UV luminosity of \mbox{T CrB} dropped by a factor of $\sim 3-4$
with respect to the average UV luminosity in 1981--85,
which explains the absence of the additional hot source
in the 1990--94 light curves.
The available data suggest the asymmetry is best visible 
in the 1981--85 $V$ light curve, thus an interpretation
in terms of asymmetric brightness distribution in 
the accretion disk seems plausible.

Such interpretation is also supported by observations of 
accretion disks in binary systems. Quiescent light curves
of dwarf novae show characteristic orbital hump, observed
during approximately one--half of the cycle, 
due to the presence of the hot spot (e.g. Warner 1995,
and references therein). Studies of the disk--accreting Algol--type
systems show that the trailing side of the disk
(where the gas stream adds to the disk) is brighter,
while the leading edge is usually more extended
(Batten 1989, and references therein).

Applying the results of Lubov \& Shu (1975) to \mbox{T CrB}
we have estimated the radius of the disk of 
$\sim 0.1 a (20 \rm R_{\odot}$), 
and the angle between the radius vector of 
the accretion stream--disk impact and the interbinary radius 
vector of $\sim 70^{\circ}$. So if there is any bright spot
in that region, its best visibility corresponds to
the orbital phase $\sim 0.8$, and it can at least qualitatively
account for the difference in heights of the photometric
maxima observed in \mbox{T CrB}. Such bright stream--disk impact
region can be also responsible for the erratic light changes
and the flickering variability,
which are apparently correlated with the average UV fluxes,
and the brightness of the additional optical continuum source.
A detailed modeling of that effect is however 
very complicated, and beyond the
scope of this paper.

\section{Summary and concluding remarks}

Based on constraints from the orbital solution for the M giant
and the amplitude of ellipsoidal light changes, and imposing
additional limits on the components masses ($M_{\rm g} \ga
0.6 \rm M_{\odot}$; $M_{\rm h} \la 1.44  \rm M_{\odot}$),
we narrow down the range of permissible values for the \mbox{T CrB} system
parameters (Table 2).
Contrary to all previous studies, our analysis shows 
that the mass ratio of \mbox{T CrB} $q \equiv M_{\rm g}/M_{\rm h} \approx 0.6$
indicating a low mass binary system, with the stellar masses 
$M_{\rm g} \sim 0.7 \rm M_{\odot}$,
and $M_{\rm h} \sim 1.2 \rm M_{\odot}$.
Our analysis also suggests that the binary orbit is circular,
and the giant seems to rotate synchronously with the orbit,
in agreement with the theoretical predictions for 
a binary with a Roche lobe--filling M giant (Zahn 1977).
Our result for the masses of the system components
solves practically all basic controversies about the nature of
the hot component and the physical causes of its eruptions.
The thermonuclear runaway in a massive white dwarf
as proposed by SCG is fully compatible with {\it all}
observational facts.

The mass ratio of \mbox{T CrB}, $q \sim 0.6$,  
is also in better agreement with the theory
of binary evolution than is the previously accepted $q \sim 1.3$.
Paczy{\'n}ski (1965b) showed that semidetached binaries
with red giant primaries can be dynamically unstable,
and recent publications demonstrate that large mass ratios
$q \ga 0.8$ are always unstable (Webbink 1988; Pastetter \& Ritter
1989). Except for the two nova--like eruptions in 1866 and 1946,
\mbox{T CrB} does not manifest any dramatic activity
that would indicate dynamically unstable mass transfer.
The erratic activity discussed in Sec. 3.1 is at similar level
as in RS Oph, a sister recurrent nova system with a massive
$\sim 1.2 \rm M_{\odot}$ white dwarf, and a low mass $\sim 0.5 
\rm M_{\odot}$ M giant companion (Shore et al. 1996; Dobrzycka \& Kenyon 1994;
Dobrzycka et al. 1996b).
SCG estimate $\dot M_{acc} \sim 
2.5 \times 10^{-8} \rm M_{\odot} yr^{-1}$ for
\mbox{T CrB}, similar to $\dot M$ derived by Dobrzycka et al. for RS Oph,
which is several orders of magnitude lower than $\dot M \sim 10^{-2}
- 10^{-3} \rm M_{\odot} yr^{-1}$ expected in the state of runaway 
mass transfer (Webbink 1988).
Moreover, according to
the theory of symbiotic binary formation and evolution 
under suitable conditions
low--mass systems containing massive white dwarfs,
although relatively rare,
may survive as symbiotic stars for a very long time
in a Roche lobe--filling state (Webbink 1988). 
\mbox{T CrB} is undoubtedly one of such systems.

We have also discussed possible causes for the asymmetry
in the visual light curve of \mbox{T CrB}. Although we cannot
propose any definitive interpretation, the most promising
is asymmetric brightness distribution in the accretion
disk surrounding the white dwarf. 
To make significant progress we need not only better observations, 
especially
in the infrared, but also improvements in the light curve analysis
and spectroscopic modeling,
including for instance implementation of M giant atmospheres option,
or line profile simulations to model both velocity field
variation across the stellar disk, and the weighted effects
of brightness asymmetries.

\section*{Acknowledgments}

This project was partially sponsored by KBN Research Grants
No. 2 P304 007 06, and No. 2 P03D 021 12.

\bsp

\label{lastpage}

\end{document}